\documentclass[twocolumn,showpacs,preprintnumbers,amsmath,amssymb,lettersize]{revtex4}

\usepackage{graphicx}
\usepackage{dcolumn}
\usepackage{bm}

\usepackage{ulem} 
\usepackage[usenames]{color}

\begin{document}

\preprint{}

\title{ 
Formation spectra of light kaonic nuclei by in-flight (${\bar K},N$) reactions with a chiral unitary amplitude}

\author{J. Yamagata-Sekihara$^1$, D. Jido$^1$, H. Nagahiro$^{2,3}$}
\author{S. Hirenzaki$^2$}%
\affiliation{$^1$Yukawa Institute for Theoretical Physics, Kyoto University, Kyoto 606-8502, Japan\\
$^2$Department of Physics, Nara Women's University, Nara 630-8506, Japan\\
$^3$Research Center for Nuclear Physics (RCNP), Osaka University, Ibaraki, Osaka 567-0047, Japan
}%

\date{\today}

\begin{abstract}
We study theoretically the in-flight ($K^-,N$) reactions for the formation of light kaonic nuclear systems to get deeper physical insights on the spectra, and to investigate the formation spectra of the reaction which will be observed at new facilities like J-PARC.
We show the expected spectra for the formation of the $K^-pp, K^-pn$, $K^-nn$ and $K^-$-$^{11}$B systems which are accessible by the ($K^-,N$) experiments.
By considering the conversion part of the Green's function, we show the missing mass spectra of the ($K^-,N$) reactions in coincident with the particle emissions due to ${\bar K}$ absorption.
To calculate the cross sections, we use the so-called $T\rho$ approximation to evaluate the optical potential.
As for the amplitude $T$, we adopt the chiral unitary amplitude of ${\bar K}N$ channel in vacuum for simplicity.
The effects of the $p$-wave optical potential of $\Sigma$(1385) channel and the contributions from ${\bar K^0}$ mixing in $^3$He($K^-,n$) reaction are also evaluated numerically.
We also study the behavior of the poles of kaon Green's function in nuclear matter.
We conclude that $^3$He($K^-,n$) and $^3$He($K^-,p$) reaction spectra in coincident with the $\pi\Sigma$ emission may show the structure in the kaon bound region indicating the existence of the unstable kaonic nuclear states.
As for the $^{12}$C($K^-,p$) spectra with the $\pi\Sigma$ emission, we may also observe the structure in the bound region, however, we need to evaluate the medium effects carefully for larger nuclei.
\end{abstract}

\maketitle
\section{\label{sec:intro}Introduction}
Kaon-nucleus bound systems, such as kaonic atoms and kaonic nuclei, have been studies to obtain information on the kaon properties in the nuclear medium {or} kaon-nucleon interaction{s} at finite density{}.
This information is important as a basic piece 
{of strangeness in nuclei}.
It is also important, for example, for heavy ion physics and astro-nuclear physics to determine the constraints on kaon condensation in high-density matter.

The $K^-$-nucleus interaction has been studied so far based on the experimental data of lightly bound kaonic atoms obtained by the X-ray spectroscopy~\cite{batty97}. 
%
{It is known that interesting features of kaon-nucleus bound systems 
are tied to the properties of kaons in nuclei which are strongly 
influenced by the change undergone by the $\Lambda(1405)$ 
in the nuclear medium, because the $\Lambda(1405)$ is a resonance 
state just below the kaon-nucleon threshold.
In fact, there are studies of kaonic atoms carried out by 
modifying the properties of the $\Lambda(1405)$ 
in the nuclear medium \cite{alberg76,wei,miz}. 
These works reproduce the properties of specific kaonic atoms reasonably well.}
The theoretical studies on kaon-nucleus optical potential based on the chiral unitary model were also reported~\cite{hirenzaki00,baca00},
%
{and our understanding of the properties of $\Lambda(1405)$ has been 
developed}
\cite{Jido:2003cb,Magas:2005vu,Hyodo:2008xr,Jido:2009jf}, recently.
{It is expected that, in near future, one completes
the study on the kaon properties  for lightly bound kaonic 
atoms  accessible by the kaonic X-ray technique.}
{The frontier of this subject is}
 in higher density region which will be studied by deeply bound kaonic atoms~\cite{gata07prc} and kaonic nuclei.

In recent years, there have been many researches in the studies of kaonic nuclear states, which are kaon-nucleus bound systems by the strong interaction inside the nucleus.
{From the theoretical study of the structure of the kaonic nuclear states 
using ${\bar K}$-nucleus optical potentials obtained in the phenomenological approach~\cite{Friedman99,friedman99} and the chiral unitary model~\cite{hirenzaki00}, it has turned out that kaonic nuclear states have large decay widths of the order of several tens of MeV.}
The kaonic nuclear states for lighter nuclei~\cite{Akaishi02,Dote:2007rk, Dote:2008hw,ikeda07,shevchenko07} and baryon resonances with two kaons~\cite{double kaon,MartinezTorres:2008kh}
 were also studied theoretically.

Common achievement of the theoretical study on the $K^{-}pp$
system~\cite{Akaishi02,Dote:2007rk,Dote:2008hw,ikeda07,shevchenko07}
is that the $K^{-}pp$ system forms a bound state with a rather large  
width.
But the binding energy and width do not reach agreement within the
theoretical calculations yet.
In the experimental side, an indication of the $K^-pp$ bound state
was observed by the FINUDA experiment \cite{finuda}.
However, the critical analyses of the latest data were reported
by Oset, Toki and their collaborators~\cite{oset06,magas06},
who claimed that the origins of the structure seen in the experimental
spectra~\cite{finuda} can be explained by the well-known processes.
The detailed analysis was also performed to understand fully
the FINUDA data~\cite{finuda06}.
Thus, the studies of the kaonic nuclear structure still  have some
sources of controversy and we need further studies.

As for the formation and observation of the kaonic nuclear states
by direct reactions, in-flight (${\bar K},N$) reactions were proposed
and performed by Kishimoto and his collaborators~ 
\cite{Kishimoto99,Kishimoto03}.
The theoretical calculations of the energy spectra
of the in-flight (${\bar K},N$) reaction were performed
in the effective number approach~\cite{gata05} and in the Green's  
function
method~\cite{arima02,gata06,Koike:2007iz}.
In Ref.~\cite{gata05}, the same theoretical technique as
Ref.~\cite{hirenzaki91} was used, which was successfully applied
for the deeply bound pionic atom formation reactions~\cite{tok,gilg00}.
In Ref.~\cite{gata06}, clear peak structure indicating the kaonic
bound states has not been found in the formation spectra
and it has turned out that it is difficult to observe
clear signals for kaonic nuclear states solely with missing-mass  
spectroscopy
due to the large $\bar K$ absorption width.
The theoretical prediction that there are no clear peaks for
the $K^{-}$-$^{11}$B bound system in the formation spectra
is consistent with the data in Ref.~\cite{kishimoto_final}, 
{which observed missing
mass spectra of the $(K^-,N)$ reactions with 1 GeV/c kaon.
Based on the data of this experiment, Ref.~\cite{kishimoto_final} extracted
the potential depth of the kaon in nuclei by fitting the formation
spectra in the Green's function method with simple background
estimation and found that the depth was as deep as
$-160$ to $-190$ MeV.}
As a new experiment, the $^3$He($K^-,n$) reaction for
the formation of the $K^-pp$ system is proposed~\cite{J-PARC}
in the new facility J-PARC. We may be able to have new experimental
information in the coming experiments in near future which will make us
possible to reach the final conclusion for the properties and/or  
existence
of the controversial kaonic nuclear states.

{As described above, there are several controversies about properties of the light kaonic nuclei.
In this exploratory level, the purpose of this article is to show the formation spectra of various light kaonic nuclei which are accessible by the (${\bar K},N$) reactions, namely $K^-pp$, $K^-pn$, $K^-nn$, ${\bar K^0}pn$, ${\bar K^0}nn$ and $K^-$-$^{11}$B systems, in order to compare formation spectra observed in forthcoming experiments. 
For this purpose, we study 
{semi-}exclusive ($K^-,N$) spectra observed
in coincidence detection of the final nucleon with the particles 
emitted by ${\bar K}$ absorption into nuclei. This is one of the main novelties 
of this work. 

To calculate the
{semi-}exclusive ($K^-,N$) spectra, we exploit 
the Green's function method with a simple $\bar K$ optical potential
obtained by a low-density $T\rho$ approximation.
The optical potential is given by the $\bar KN$
scattering amplitude $T$ and the nuclear density distribution $\rho(r)$.
The $\bar KN$ scattering amplitude is evaluated in 
coupled channels of $\bar KN$, $\pi \Sigma$, $\pi \Lambda$, $\eta \Sigma$,
$\eta \Lambda$ and $K \Xi$ based on the chiral unitary approach and the $\Lambda(1405)$ is dynamically 
generated. This potential automatically includes one-body $\bar K$ absorption 
to the $\pi \Sigma$ and $\pi \Lambda$ channels. Thus we can calculate
the semi-exclusive processes of the kaonic nuclei decays including 
$\bar KN \to \pi \Sigma$ and $\pi \Lambda$. 
We assume a simple two parameter Fermi function for the nuclear density $\rho(r)$,
which should be calculated dynamically by few-body treatments though.
As we will see later, even with such a simple estimation of the optical potential,
the obtained potential produces a similar bound state structure with
one calculated by an elaborated few-body formulation. Thus, we expect
to obtain very similar formation spectra with the few-body formulation.

In the present article, we also discuss 
the effects of the $p$-wave optical potential of the $\Sigma (1385)$ channel and the contributions from ${\bar K^0}$ mixing in $^3$He($K^-,n$) reaction.
As for the $\bar KN$ amplitude $T$, to see the effects of the medium modifications, we compare the spectra calculated with the in-vacuum amplitude and those obtained with the in-medium kaon self-energy evaluated beyond the low density approximation based on the chiral unitary model.  }

\begin{table}[htbp]
\renewcommand{\arraystretch}{1.2} 
\begin{center}
\caption{\label{tab:state} 
Possible ${\bar K}NN$ systems and accessible ($K^-,N$) reactions.
Optical potentials used in the calculation of reaction are also listed. }
\begin{tabular}{cccc}
System&Formation reaction&Optical potential \\ \hline\hline 

$K^-pp$&$^3$He($K^-,n$)&Eq.~(\ref{V:Kpp})\\ \hline
$K^-pn$&$^3$He($K^-,p$)&Eq.~(\ref{V:Kpn})\\ 
&$t$($K^-,n$)&Eq.~(\ref{V:Kpn})\\ \hline
$K^-nn$&$t$($K^-,p$)&Eq.~(\ref{V:Knn})\\ \hline
${\bar K^0}pn$&$^3$He($K^-,n$)&Eq.~(\ref{V:Kpn})\\ \hline
${\bar K^0}nn$&$t$($K^-,n$)&Eq.~(\ref{V:Kpp})\\ \hline
\end{tabular} 
\end{center}
\end{table}

In Sec.~\ref{sec2}, we describe the theoretical models for the studies of the formation of the ${\bar K}NN$ systems and the ${\bar K}$ optical potential with the chiral unitary ${\bar K}N$ amplitude.
In Sec.~\ref{sec3}, we study the behavior of the poles of kaon Green's function in nuclear (proton) matter.
Numerical results of the formation of the ${\bar K}NN$ and $K^-$-$^{11}$B systems are presented and discussed in Sec.~\ref{sec4}.
We give conclusions of this article in Sec.~\ref{sec5}.

\section{\label{sec2}Formalism}
To investigate the formation of the light kaonic nuclei such as ${\bar K}NN$ systems, we apply basically the same theoretical framework as our previous publications~\cite{gata06,gata07prc}.
We solve the Klein-Gordon equation to obtain the eigenenergies of the states and apply the Green's function method~\cite{morimatsu85} to calculate the reaction spectra.
In this section, we explain our theoretical model for studying light kaonic nuclear systems in detail.

We consider the ($K^-,N$) reactions as the production reaction of light kaonic nuclei 
{with a 600 MeV incident energy of $K^{-}$ in the lab.\ frame.}
In this reaction, the incident kaon kicks out 
{one of the nucleons in}
the target nucleus and the kaon-nucleus system is produced. 
{The emitted nucleon is observed in the forward direction to make the momentum transfer smaller.}
Because the incident kaon has large momentum, the impulse approximation may be good to evaluate the formation spectrum.
We
use the local density approximation to evaluate the medium effects on kaon and
do not consider the multi step process that the kaon hits many nucleons in daughter nucleus.

{The energy conservation of this reaction in the lab.\ frame is written as $E_{K^-}+M_{A}=E_{A-1\otimes K^-}+E_N$, where $E_{K^-}$ and $M_{A}$ indicate the energy of the initial kaon and the mass of the target nucleus, and $E_{A-1\otimes K^-}$ and $E_N$ are the energies of the kaon-nucleus systems and the emitted nucleon.
The momentum of the kaon-nucleus system, equivalently the transferred momentum,  is indicated as ${\bm q}$.
The momentum conservation is written as ${\bm p}_{K^-}={\bm q}+{\bm p}_N$, where ${\bm p}_{K^-}$ and ${\bm p}_N$ indicate the momenta of the initial kaon and the emitted nucleon, respectively. In the present reaction, the transferred momentum is about 200 MeV/c in the lab. frame. 
We assume that the recoil kinetic energy of the kaon-nucleus system, 
$T=|{\bm q}|^2/(2M_{A-1\otimes K^-})$, is small and negligible.
Writing the target nucleus mass as $M_{A} = M_{A-1} + M_{N} - S_{N}$ with  
the one nucleon separation energy $S_N$, 
and the kaon energy measured from the in-vacuum kaon mass (binding energy) as $E = M_{A-1\otimes K^-}-M_{A-1}-M_{K^-}$, we obtain 
$T_N=T_{K^-}-S_N-E$, where $T_{K^-}$ is the incident kaon kinetic energy, and $T_N$ the emitted nucleon kinetic energy.
 }

In the impulse approximation, the expected spectra of the ($K^-,N$) reactions $\displaystyle{\Bigl(\frac{d^2\sigma}{d\Omega dE_N}\Bigr)}$ are evaluated by the nuclear response function $S(E)$ and elementary cross section $\displaystyle{\Bigl(\frac{d\sigma}{d\Omega}\Bigr)^{\rm ele}}$ as,
\begin{equation}
\label{eq:6}
\displaystyle{\Bigl(\frac{d^2\sigma}{d\Omega dE_N}\Bigr)=\Bigl(\frac{d\sigma}{d\Omega}\Bigr)^{\rm ele}\times S(E)}~~.
\end{equation}

We use the Green's function method~\cite{morimatsu85} to calculate the response function of the light kaonic nuclei in the ($K^-,N$) reactions. The details of the application of the Green's function method are found in Refs.~\cite{gata06,hayano99,klingl99,jido02,nagahiro05,gata07prc}. 
The calculation of the nuclear response function with the complex potential is formulated by Morimatsu and Yazaki~\cite{morimatsu85} as 
\begin{equation}
\label{S(E)}
S(E)=-\frac{1}{\pi} {\rm Im}\sum_f \int d{\bm r}d{\bm r'} \tau^\dagger_f G(E; {\bm r},{\bm r'}) \tau_f,
\end{equation}
\noindent
where the summation is taken over all possible final states.
$G(E; {\bm r},{\bm r'})$ is the Green's function of kaon interacting in the nucleus and defined as,
\begin{equation}
\label{Gfunc}
G(E; {\bm r},{\bm r'})=\langle\alpha |\phi_{K}({\bm r})\displaystyle{\frac{1}{E-H_K+i\epsilon}}\phi^+_K({\bm r'})|\alpha\rangle~~,
\end{equation}
where $\alpha$ indicates the proton hole state and $H_K$ indicates the Hamiltonian of the kaon-nucleus system.
The amplitude $\tau_f$ denotes the transition of the incident particle (${\bar K}$) to the nucleon-hole and the outgoing nucleon, involving the nucleon-hole wavefunction $\psi_{j_N}$ and the distorted waves $\chi_i$ and $\chi_f$, of the projectile and ejectile.
By taking the appropriate spin sum, the amplitude $\tau_f$ can be written as,
\begin{equation}
\label{tau}
\tau_f({\bm r})=\chi_f^*({\bm r})\xi_{1/2,m_s}^*[Y_{l_{\bar K}}^*(\hat{\bm r})\otimes \psi_{j_N}({\bm r})]_{JM}\chi_i({\bm r})~,~
\end{equation}
\noindent
with the meson angular wavefunction $Y_{l_{\bar K}}(\hat{\bm r})$ and the spin wavefunction $\xi_{1/2,m_s}$ of the ejectile.
We assume the harmonic oscillator wavefunctions for $\psi_{j_N}$ with the range parameter determined by the r.m.s radii of the target nuclei.

The
{semi-}exclusive spectra can be calculated by decomposing the response function~(\ref{S(E)}) into the escape and conversion parts: $S=S_{\rm esc}+S_{\rm con}$.
This decomposition can be done exactly by
\begin{eqnarray}
&&S_{\rm esc}(E)\nonumber\\
&&=-\displaystyle{\frac{1}{\pi}}\sum_f \int d{\bm r}d{\bm r'} \tau_f^\dagger (1+G^\dagger V_{\rm opt}^\dagger){\rm Im}G_0(1+V_{\rm opt}G)\tau_f\nonumber\\
&&S_{\rm con}(E)=-\displaystyle{\frac{1}{\pi}}\sum_f  \int d{\bm r}d{\bm r'}\tau_f^\dagger G^\dagger {\rm Im}V_{\rm opt}G\tau_f~~.
\label{eq:9}
\end{eqnarray}
\noindent
where $V_{\rm opt}$ is the kaon-nucleus optical potential given in Hamiltonian.
The conversion part is known to express the contributions of the kaon absorption to the (${\bar K},N$) spectra~\cite{morimatsu85}.
We can further decompose the conversion part of the response function (and clearly the (${\bar K},N$) spectra) into different particle emission channels as,
\begin{equation}
\label{Scon_con}
S_{\rm con}(E)=\sum_j S_{\rm con}^{(j)}(E)~~,
\end{equation}
where $j$ indicates the decay channel. In this work we consider $j={\bar K}N,~\pi\Sigma,~\pi\Lambda,~\eta\Lambda$ channels. The separated spectrum to each channel can be compared with the experimental spectrum observed in coincident measurements of emitted particles due to the kaon absorption.

To calculate the Green's function $G(E; {\bm r}, {\bm r'})$, we use multipole expansion,
\begin{equation}
\label{Eq:A65}
G(E; {\bm r}, {\bm r'})=\sum_{Lm}Y^*_{Lm}(\Omega') Y_{Lm}(\Omega) G^L (E; r, r')~~,
\end{equation}
with the solution $G^L (E; r, r')$ of the radial equation,
\begin{eqnarray}
&&(\frac{d^2}{dr^2}+\frac{2}{r}\frac{d}{dr}-\frac{L(L+1)}{r^2}-2\mu V_{\rm opt}\nonumber\\
&&~~~~~~~~~~~~~~~~~-2\mu V_{\rm coul}+2\mu E)G^L(E;r,r')\nonumber\\
&&=-\frac{2\mu}{r^2}\delta(r-r')~~,
\label{Eq:A88}
\end{eqnarray}
where $\mu$ is the reduced mass of the $\bar K$ and nucleus system and $V_{\rm coul}$ is the Coulomb potential with a finite nuclear size.
The strength of the energy dependent potential used in this article is also evaluated using this kaon energy.
{To solve the Eq.~(\ref{Eq:A88}), the kaon energy $E$ is regarded as an external parameter.}
The solutions of the radial equation can be expressed as
{\cite{newton}},
\begin{equation}
G^L(E;r,r')=\displaystyle{\frac{-2\mu}{(r')^2\Delta(u_L(r'),v_L(r'))}} u_L(r_<)v_L(r_>)~~,
\label{Eq:A108}
\end{equation}
with the regular solution $u_L(r)$ at the origin, the outgoing solution $v_L(r)$ of the homogeneous equation of Eq.~(\ref{Eq:A88}) for the fixed values of $E$
{~and the wronskian $\Delta(f,g)=fg^\prime - f^\prime g$ where
$f^\prime$ denotes the derivative of $f$ with respect to $r^\prime$.}
Here, we have introduced the symbol $r_>$ ($r_<$) to indicate the larger  
(smaller) radial coordinate variable of $r$ and $r'$
{.}

For the values of the nucleon separation energies $S_N$, 
we use the binding energies of $^{3}$He and $t$,
7.72 MeV and 8.48~MeV, for the 
$^3$He$(K^-,n)$ and $t(K^-,p)$ reactions, respectively, since 
the residual two-nucleon systems are not bound, while, for the
$^3$He($K^-,p)$ reaction, we use the proton separation energy 
of $^{3}$He, which is the energy difference of $p+d$ and $^{3}$He
systems. 
For the $^{12}$C($K^-,p$) reaction,
we use the measured values of the $s_{1/2}$ and $p_{3/2}$ 
proton separation energies of $^{12}$C given in the Table of Isotopes~\cite{isotope}.

{
Since we are interested in the calculation of the formation spectra of the kaonic 
nuclei, we use the one-body Green function of the in-medium kaon, in which
the interactions of the kaon and nucleus are expressed by the optical potential
$V_{\rm opt}$. The optical potential is evaluated simply by the low density 
approximation, in which the optical potential is given by the forward 
scattering amplitude of the kaon and nucleon in vacuum. 
For the description of the 
$\bar KN$ scattering amplitude, we exploit the chiral unitary approach, in which
the $\bar KN$ interactions are summed up in a non-perturbative way 
in the coupled-channels formulation and the $\Lambda(1405)$ is dynamically
generated in the $\bar KN$ scattering. Since the iterative 
$\bar K$ interactions with one nucleon in the nucleus is already 
taken into account in the construction of the scattering amplitude, we include
a correction factor $(A-1)/A$ {for the nuclear density} to avoid double-counting of the iteration
with the same nucleon. Thus, the optical potential is given in the 
low density approximation by
}
\begin{equation}
   V_{\rm opt} = - \frac{1}{2\mu} \frac{A-1}{A} T(E) \rho \label{eq:Trho}
\end{equation}
{where $\rho$ is the nuclear density and $T(E)$ is the $\bar KN$ scattering amplitude calculated in vacuum. The $s$-wave $\bar KN$ amplitude 
is taken from Ref.~\cite{Oset:2001cn}. 
}
The importance of the energy dependence of the interactions is emphasized in Refs.~\cite{gata06,gata07} especially for the imaginary part.
We include the full energy dependence of the interactions in microscopic way~\cite{Oset:2001cn} both for real and imaginary parts in this article.

The optical potentials between ${\bar K}$ and $NN$ systems, then, can be written as,
\begin{equation}
\label{eq:Vopt}
V^{{\bar K}NN}_{\rm opt}(r,E)=\displaystyle{\frac{1}{2}}V_s^{{\bar K}NN}(E)\rho_{NN}(r)~~,
\end{equation}
where $\rho_{NN}$ indicates the density profile of the two nucleon systems and $V_s^{{\bar K}NN}$ the $s$-wave potential strength 
{given} 
by the $s$-wave chiral unitary ${\bar K}N$ amplitude $T_s$ as,
\begin{eqnarray}
\label{V:Kpp}
V^{K^-pp}_s&=&V_s^{{\bar K^0}nn}=\displaystyle{\frac{1}{2\mu}\frac{1}{2}}(T_s(I=0)+T_s(I=1))~~,\\
V^{K^-pn}_s&=&V^{{\bar K^0}pn}_s\nonumber\\
&=&\displaystyle{\frac{1}{2\mu}\frac{1}{2}}\left(\displaystyle{\frac{T_s(I=0)+T_s(I=1)}{2}}+T_s(I=1)\right)~~, \nonumber\\
&~&
\label{V:Kpn}
\\
\label{V:Knn}
V^{K^-nn}_s&=&\displaystyle{\frac{1}{2\mu}}T_s(I=1)~~.
\end{eqnarray}
Here, $I$ indicates the isospin of ${\bar K}N$, and $\mu$ the reduced mass of the kaon and the $NN$ system.

The optical potential for the $K^-$-$^{11}$B system is obtained in the similar way:
{
\begin{eqnarray}
V_{\rm opt}^{^{11}{\rm B}-K^-}&=&\frac{1}{2\mu}\rho_{^{11}{\rm B}}(r)\nonumber\\
&\times&\left(\frac{4}{11}\frac{T_s(I=0)+T_s(I=1)}{2}+\frac{5}{11}T_s(I=1)\right).\nonumber\\
\label{eq:11B}
\end{eqnarray}
The first and second terms represent the $K^{-}p$  and $K^{-}n$ interactions,
respectively. Here we have assumed that the density distributions for the proton and neutron have the same radius and diffuseness parameters. }

In this article, we also study the contribution of the $p$-wave optical potential which includes $\Sigma(1385)$ effects.
We rewrite the Eq.~(\ref{eq:Vopt}) with the $p$-wave potential as,
\begin{equation}
\label{eq:Vopt_p}
V^{{\bar K}NN}_{\rm opt}(r,E)=\displaystyle{\frac{1}{2}}(V_s^{{\bar K}NN}\rho_{NN}-\vec\nabla\cdot V_p^{{\bar K}NN}\rho_{NN}\vec\nabla)~~,
\end{equation}
\noindent
where $V_p^{{\bar K}NN}$ indicates the $p$-wave potential strength defined by the $p$-wave ${\bar K}N$ amplitude $T_p$ by the same isospin combinations as in Eqs.~(\ref{V:Kpp})-(\ref{V:Knn}).
$T_p$ is defined as the first term of Eq.~(18) in Ref.~\cite{Jido:2002zk} including factor 3.

In our model, we assume the density profile of two nucleons as the form of the two parameter Fermi function
with the radius $R$ and diffuseness $a$. 
Here we take $R_{pp}=R_{nn}=1.01$ fm and $a_{pp}=a_{nn}=0.50$ fm 
for the $pp$ and $nn$ systems, which provide $\rho_{pp}(0)=\rho_{nn}(0)=0.117$ fm$^{-3}$, and $R_{pn}=1.40$ fm and $a_{pn}=0.51$ {fm} for 
the $pn$ system, which is a similar with the deuteron density distribution. 
As we will see later in Sec.~\ref{sec4}, we will try other 
radius parameters and will find that  
the calculated spectra is insensitive to the change of 
the parameter.
We do not specify the isospin of the $NN$ systems here.
We also use the two parameter Fermi density distribution 
for $^{11}$B as in Ref.~\cite{gata06} with parameters $R=2.23$ fm and $a=0.44$ fm.

In the following sections, we compare the formation spectra calculated with the optical potential given in Eq.~(\ref{eq:Trho}) with those obtained with the
in-medium kaon self-energy calculated beyond the low density approximation 
in the chiral unitary approach done in Ref.~\cite{ose}. Since the latter potential
has higher-order medium effects beyond two-nucleon processes, it is not 
appropriate for the three-body system. But it may be interesting to know  
the medium effects beyond the low density approximation and 
the uncertainties of our theoretical framework.

To calculate the conversion part of different particle emission channels~(\ref{Scon_con}), we consider the optical theorem satisfied by the chiral unitary amplitude $T$ as,
\begin{equation}
\label{chiralT_con}
{\rm Im}~T_{{\bar K}N\rightarrow {\bar K}N}=\sum_j T_{{\bar K}N\rightarrow j}\sigma_j T_{j \rightarrow {\bar K}N}^*~~,
\end{equation}
\noindent
where $j$ indicates the intermediate state considered to calculate the ${\bar K}N\rightarrow{\bar K}N$ amplitude $T_{{\bar K}N\rightarrow {\bar K}N}$ by the chiral unitary model and means $j={\bar K}N,~\pi\Sigma,~\pi\Lambda,~\eta\Lambda,~\eta\Sigma$, and $\kappa\Xi$, and $\sigma_{j}$ is
the two-body phase space of the intermediate state $j$.
We can interpret each term in the right hand side as the contribution of the each intermediate channel to the absorptive part of ${\bar K}$ optical potential in the $T\rho$ approximation.

In the present calculations, the final state interaction effects to the emitted particles from ${\bar K}$ absorption are not considered, which should be included for more quantitative results.
And the only one-body absorption processes are evaluated here and the two-body absorption processes such as ${\bar K}NN\rightarrow N\Lambda$ are not included in this calculation.

For the simple estimations of kaon bound states for the light kaonic nuclear systems, we solve the Klein-Gordon equation with local density approximation assuming two nucleon density distribution. 
And we solved numerically as in Refs.~\cite{gata05,gata06,gata07prc}, following the method of Oset and Salcedo~\cite{oset85}.
Since we adopt the theoretical optical potential in Eq.~(\ref{eq:Vopt}) with the complex energy dependence, we solve the Klein-Gordon equation iteratively to obtain selfconsistent solution for complex energy $E$.

\section{Kaon pole in nuclear matter \label{sec3}}
In this section, we study the behavior of the kaon poles in the nuclear matter before showing the (${\bar K},N$) spectra.
We use the $s$-wave optical potential in Eq.~(\ref{eq:Vopt}) and obtain the pole energies by solving the Klein-Gordon equation in nuclear matter,
\begin{equation}
\label{eq:pole}
\omega ^2-m_K^2-2m_KV_s^{{\bar K}NN}(\omega)\rho=0~~,
\end{equation}
\noindent
where the $m_K$ is the kaon mass, the $\rho$ is the nucleon density of the matter and $\omega$ is written as $\omega=E+m_K$.
We neglect the kaon kinetic energy and the $p$-wave optical potential terms by postulating the small momentum of the bound kaon.

\begin{figure}[htpd]
\includegraphics[width=8.6cm,height=6.0cm]{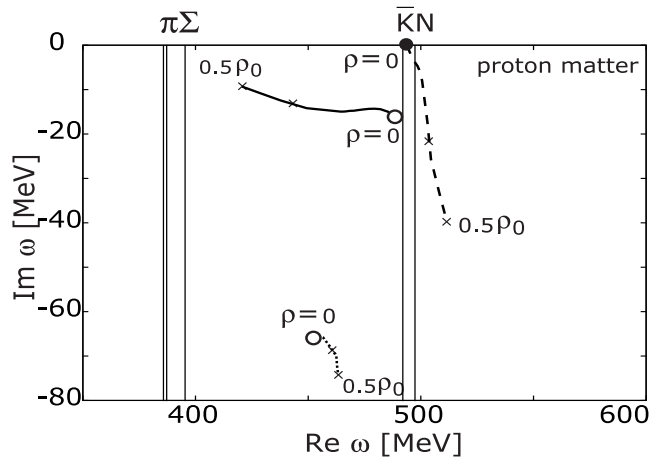}
\caption{\label{fig:pole}
Trajectories of the poles in the complex $\omega$ plane of $K^-$ propagator in proton matter obtained by solving the Klein-Gordon equation Eq.~(\ref{eq:pole}) for various proton density $\rho$.
In vacuum ($\rho=0$), only single pole appears at the energy of the free kaon mass.
Crosses correspond to the step size of density $\delta \rho=0.25\rho_0$ with $\rho_0=0.17$ fm$^{-3}$.
The vertical lines indicate the threshold energies of neutral ${\bar K}N$ and $\pi\Sigma$ channels.}
\end{figure}

The calculated pole positions of $K^-$ in the proton matter, where $\rho$ indicates the proton density and no neutron is in the matter, are shown in Fig.~\ref{fig:pole} in the complex energy plane.

As we can see from the figure, we have only one pole at $\rho=0$ at $\omega=m_K$ corresponding to the free kaon.
At finite density $\rho>0$, there appear two other poles which correspond to the $\Lambda(1405)$ resonance of the ${\bar K}N$ system.
As pointed out in Ref.~\cite{Jido:2003cb}, both of the two poles appeared at $\rho>0$ correspond to the $\Lambda(1405)$ resonances.
As increasing the density $\rho$ from $0$, we find that the kaon pole at $\omega=m_K$ moves to the direction to higher Re $\omega$ and wider width $\Gamma=-2{\rm Im}~\omega$, as shown by the dashed line in Fig.~\ref{fig:pole}.
One of the $\Lambda(1405)$ pole shown by the solid line moves to lower Re $\omega$ region with smaller width $\Gamma$, which indicates the possible existence of the narrow peak structure in the (${\bar K},N$) spectra in finite nucleus.
In this calculation, we have only included the one-body processes for the kaon absorption, and thus the width $\Gamma$ become smaller 
in this energy region because of the threshold effects of the $\pi\Sigma$ decay.
The third pole, which starts from $\omega=(452.2,-66.0)$ MeV, moves inside the Im $\omega=-65$ -- $-80$ MeV region for $\rho=0$ -- $0.5\rho_0$ and never comes closer to Re $\omega$ axis.

From the behaviors of the poles in proton matter, we may have a schematic picture of the (${\bar K},N$) spectra.
Since the kaon pole at $\omega=m_K$ at $\rho=0$ gets larger width $\Gamma\sim 80$ MeV and larger real energy at $\rho=0.5\rho_0$, this pole is expected to contribute to the quasi-elastic energy region in the spectra with large width as a smooth slope.
The pole shown as the solid line in Fig.~\ref{fig:pole}, which is one of the $\Lambda(1405)$ poles, has smaller width at larger density and, is expected to be seen as peak structures in the reaction spectra, which could be considered as candidates of the signals of $K^-pp$ states.
As seen in Fig.~\ref{fig:pole}, the state (pole) is originated from the one of the $\Lambda(1405)$ poles in the matter and are not connected to free kaon at $\rho\rightarrow 0$ limit.
This is one of the interesting findings in our framework.
In the experimental spectra, this single pole may provide several peak structures, since in quantum mechanics we have in general more than one discrete bound level in the finite system for a single pole in the infinite system.
The third pole only moves around Im $\omega = -65$ -- $-80$ MeV ($\Gamma=130$ -- $160$ MeV) region, and thus, will not make any clear structures in the spectra.
For further qualitative analyses, one needs to calculate the optical potential beyond the low density approximation and treat the meson self-energies in a self-consistent way as done in Refs.~\cite{lut}.

\section{\label{sec4}Numerical Results}
\subsection{\label{sec4-1}Kaon optical potential and bound states for finite systems}
\begin{figure*}[htpd]
\includegraphics[width=13cm,height=9.5cm]{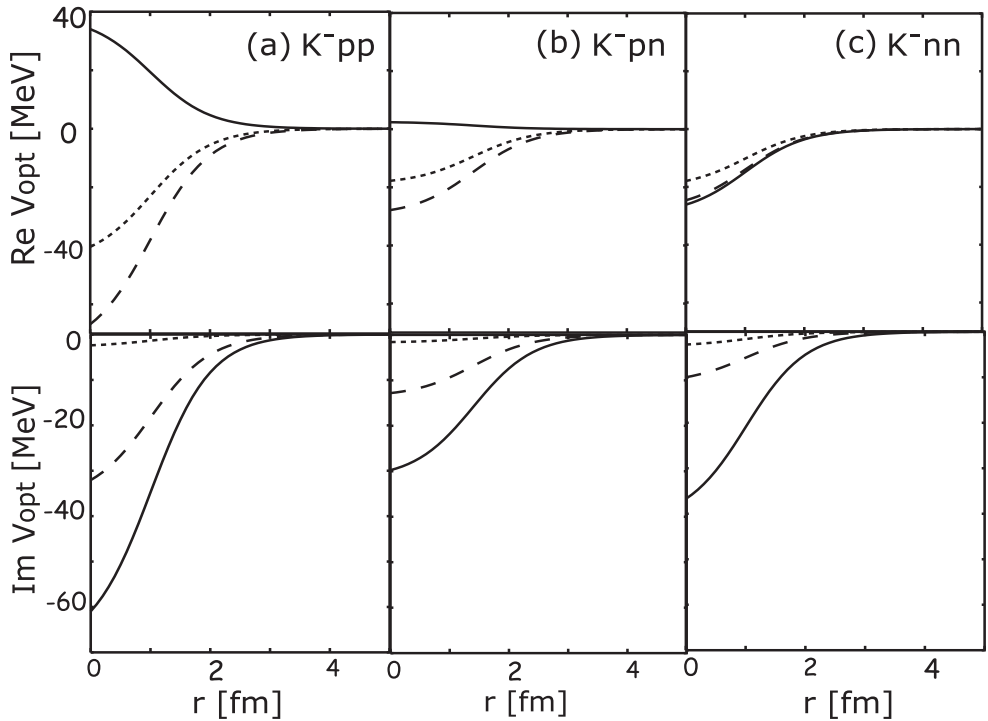}
\caption{\label{Vopt}
The $s$-wave $K^-$ optical potentials defined in Eq.~(\ref{eq:Vopt}) with the free space chiral ${\bar K}N$ amplitude $T$~\cite{Oset:2001cn} in the $T\rho$ approximation as a function of the radial coordinate $r$ for the (a)$K^--pp$, (b)$K^--pn$, and (c)$K^--nn$ systems, respectively.
The upper and lower panels show the real and imaginary parts.
The solid, dashed, dotted lines indicate the potential strength for the real kaon energies Re $E\equiv{\rm Re}~\omega-m_K=0$ MeV, $-50$ MeV, and $-100$ MeV with Im $E=0$, respectively.}
\end{figure*}

We show firstly in Fig.~\ref{Vopt} the $s$-wave optical potential defined in Eq.~(\ref{eq:Vopt}) to see the strength and the energy dependence of the microscopic potential.
The contribution of the $p$-wave potential will be discussed later.
Since we use the microscopic ${\bar K}N$ scattering amplitude of the chiral unitary model, we can consider the potentials of the $K^-$-$pp$,~$K^-$-$pn$, and $K^-$-$nn$ systems simultaneously.
The energy dependence of each potential appears theoretically in both real and imaginary parts.

As we can find easily in Fig.~\ref{Vopt}, the microscopic optical potentials have strong energy dependence and channel dependence.
For the real part of the potential shown in the upper panels in Fig.~\ref{Vopt}, we can see that the $K^-$-$pp$ potential is repulsive at the threshold energy corresponding to the negative sign of the $K^-p$ scattering length, while it shows the strong attractive nature at lower energies due to the existence of the $\Lambda(1405)$ resonance.
The real potential for $K^-$-$pn$ has rather milder energy dependence than the $K^-$-$pp$ case, and it is almost energy independent for the $K^-$-$nn$ system by reflecting there are no baryon resonance in the $s$-wave $K^-n$ system in this energy region.
As for the imaginary potential, only one-body absorption processes are included for all cases shown in Fig.~\ref{Vopt}, and thus the strength of the absorption is significantly weaker for smaller kaon energies due to the smaller phase space of $\pi\Sigma$ decay. 
As for the $p$-wave potential introduced in Eq. (\ref{eq:Vopt_p}), the strengths of the $V_p^{K^-pp}$ at the nuclear center are Re $V_p^{K^-pp}(0)\sim 2.5 \times 10^{-4}$ [MeV$^{-1}$] and Im $V_p^{K^-pp}(0)\sim - 0.8 \times 10^{-4}$ [MeV$^{-1}$] at the kaon threshold energy.

We have solved the Klein-Gordon equation with these potentials in self-consistent manner for the complex kaon energy in finite nuclear systems.
The eigenenergies obtained in the calculation are compiled in Table~\ref{tab:eigen}.
\begin{table}[htbp]
\begin{center}
\caption{\label{tab:eigen}Calculated binding energies and widths in unit of MeV for the $K^-NN$ systems using the optical potential defined in Eq.~(\ref{eq:Vopt}).
The widths are shown in parenthesis.}
\begin{tabular}{c|c|c|c}
[MeV]&$~~K^-pp~~$&$~~K^-pn~~$&$~~K^-nn~~$ \\ \hline
$1s$&23.0(41.3)&12.9(36.4)&--\\ \hline
$2s$&9.0(36.8)&--&--\\ \hline
$2p$&11.5(38.5)&8.6(35.4)&--\\ \hline
\end{tabular} 
\end{center}
\end{table}
As the consequences of the strong complex energy dependence of the optical potential, we have found interesting spectra of the bound states in $K^-pp$, where the $2s$ level is significantly close to the $2p$ level.
Because the depth of the energy dependent potential {becomes} so deep at the energy of $2p$ state in complex energy plane and the centrifugal force is considered in the Klein-Gordon equation, the kaon-nucleon subsystem has the component with $l\ne 0$.
As we can expect from the potential shape in Fig.~\ref{Vopt}, the number of bound states is different for each system.
We have also evaluated the effects of the $p$-wave optical potential and found that the variations of the binding energies and widths due to the $p$-wave potential are lesser than 10$\%$ for the states shown in Table~\ref{tab:eigen}.
In all cases considered here, the self-consistent solutions in the complex energy plane have large decay widths, and thus the pole position in the complex plane is far from the real axis.

We also notice that since we have assumed the $\rho_{NN}$ profile in this article, we should consider the results listed in Table~\ref{tab:eigen} as qualitative.
These binding energies and widths could be interesting to interpret the formation spectra calculated with the same density distributions and interactions.
Even in our simple framework, the calculated results shown in Table~\ref{tab:eigen} resemble to the results of the bound states obtained in three-body calculations with a variational approach~\cite{Dote:2008hw},
{in which they use a $\bar KN$ effective interaction derived by
the chiral unitary approach~\cite{Hyodo:2007jq}}.
Since the bound state structures are similar with each other, the formation spectra obtained in the present calculation could be like the spectra calculated with the three-body wavefunction obtained in Ref.~\cite{Dote:2008hw}.

\begin{figure}[htpd]
\includegraphics[width=8.6cm,height=10.0cm]{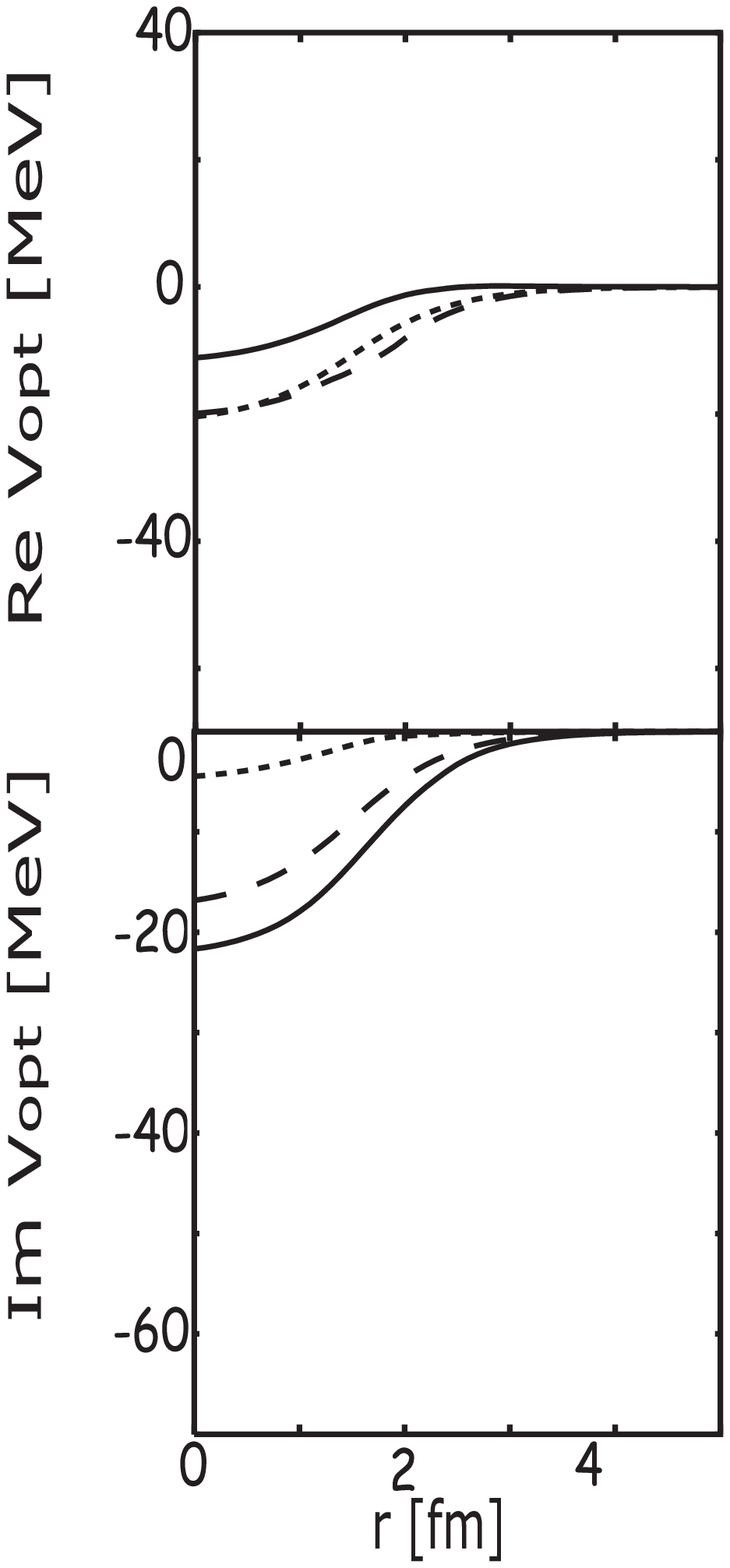}
\caption{\label{Voptmedium}
The $K^--pn$ $s$-wave optical potential of the chiral unitary approach based on the $K^-$ selfenergy in the symmetric nuclear matter \cite{ose}, as a function of the radial coordinate $r$.
The upper and lower panels show the real and imaginary parts, respectively.
The solid, dashed, and dotted lines indicate the potential strength for the kaon real energies Re $E\equiv{\rm Re}~\omega-m_K=0$ MeV, $-50$ MeV, and $-100$ MeV with Im $E=0$ MeV, respectively.}
\end{figure}

In Fig.~\ref{Voptmedium}, we show the optical potential for the $K^-$-$pn$ system evaluated by the $K^-$ selfenergy in the symmetric nuclear matter~\cite{ose} with medium modifications of the chiral unitary amplitude.
We see that the real part of the potential is attractive even at the threshold by the medium effects and has smaller energy dependence than that shown in Fig.~\ref{Vopt} (b).
The imaginary potential, which includes the multi-nucleon absorption effects, also has milder energy dependence.

\subsection{Inclusive spectra of ${\bar K}NN$ system formation}
\label{sec:4-2}
\begin{figure*}[htpd]
\includegraphics[width=13.cm,height=8.0cm]{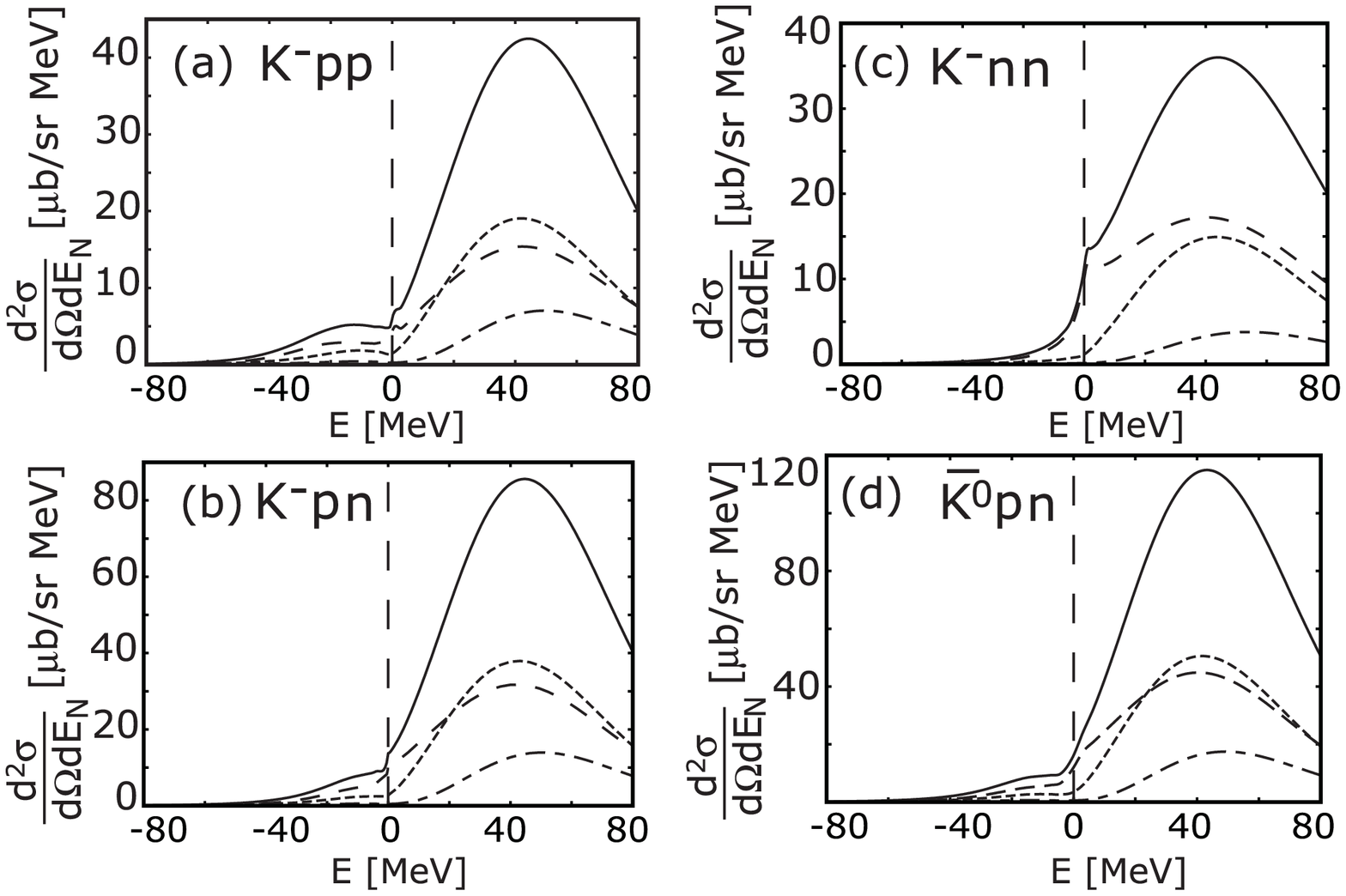}
\caption{\label{Kpptot}
Calculated spectra for the formation of (a) $K^-pp$ in $^3$He($K^-,n$), (b) $K^-pn$ in $^3$He($K^-,p$), (c) $K^-nn$ in $t$($K^-,p$), and (d) ${\bar K^0}pn$ in $^3$He($K^-,n$) reactions at $T_{K^-}=600$ MeV ($P_{K^-}=976$ MeV/c) are plotted as a function of the real kaon energy at $\theta_N^{\rm lab}=0$ (deg.) for the $s$-wave chiral unitary optical potential.
The horizontal axis $E$ indicates the real kaon energy.
Solid line shows the total spectra.
Dashed, dotted and dot-dashed lines indicate the contributions from kaon $s$, $p$, and $d$ partial waves in the final state, respectively.
The vertical dashed line indicates the kaon production threshold.
}
\end{figure*}

We show the calculated inclusive $^3$He($K^-,n$) spectra in Fig.~\ref{Kpptot} (a) for the formation of the $K^-pp$ system with the $s$-wave optical potential defined in Eq.~(\ref{eq:Vopt}).
We find that there is a certain bump structure in the subthreshold region which could be identified as the signal of the $K^-pp$ system.
Actually we can find the eigen solution of the Klein-Gordon equation in this energy region.
This pole is expected to be connected to $\Lambda(1405)$ resonance in $\rho\rightarrow 0$ limit as we have discussed in Sec.~\ref{sec3}.
Thus, it is expected that this structure can be the signal of the mixture of the $K^-pp$ and $\Lambda(1405)p$ states.
It would be interesting to study the origins of the structure in the spectra in detail as reported recently in Ref.~\cite{Jido:2008ng} for $\eta$ mesic nucleus case in connection with the $N^*(1535)$ state in nucleus.

In Figs.~\ref{Kpptot} (b) and~\ref{Kpptot} (c), we consider other ${\bar K}NN$ systems accessible by the (${\bar K},N$) reactions.
In Fig.~\ref{Kpptot} (b), we consider the $K^-pn$ system formation by $^3$He($K^-,p$) reactions.
Since the optical potential of $K^-$-$pn$ is less attractive than the $K^-$-$pp$ case in the subthreshold region as shown in Fig.~\ref{Vopt}, the bump structure in Fig.~\ref{Kpptot} (b) is smaller than that in Fig.~\ref{Kpptot} (a).
This subthreshold bump for $K^-$-$pn$ formation, however, still reflects the existence of the pole of Klein-Gordon equation in the complex energy plane and is interesting to be studied experimentally.
In Fig.~\ref{Kpptot} (c), we show the expected spectra for the formation of the $K^-nn$ system by $t$($K^-,p$) reaction.
As we can expect from the weakest $K^-$-$nn$ interaction in all systems considered here, we do not find the subthreshold structure in the spectra in this case.
We only expect to observe a possible cusp structure at the threshold which could be a reminiscent of bound states.

The calculated results of the formation spectra for the ${\bar K}NN$ systems in Figs.~\ref{Kpptot} (a)-(c) show the characteristic features indicating the differences of the structures and optical potentials for the $K^-pp$, $K^-pn$, and $K^-nn$ systems.
It would be interesting that the measurements will be performed for all three channels and compared each other to see the differences.
The signals appearing in the inclusive spectra as enhancements in the bound energy region, however, are small portions of the whole spectra in general and thus, we also consider the 
{semi-}exclusive spectra later in this section.

Here we discuss the isospin relation among the $\bar K NN$ formation spectra. 
Since we use the isospin symmetric optical potentials as given 
in Eqs.~(\ref{V:Kpp}) and (\ref{V:Kpn}) and 
we do not consider the $NN$ correlations, namely the isospin dependence
in the $NN$ subsystems, in the present calculation, the {Green's} functions are 
same for the $K^-pp$ and ${\bar K^0}nn$ systems, 
and for the $K^-pn$ and ${\bar K^0}pn$ systems, respectively. 
This means that, if there are bound states, the spectra of the bound states
are equivalent in each pair of the systems. 
But, since the elementary cross section and the distortion factor are different
in each system, the formation spectra obtained with these potentials are 
{quantitatively} different in the magnitude.  For example, we show the calculated {spectrum} for the ${\bar K^0}pn$ {system} in the $^3$He($K^-,n$) reaction in Fig.~\ref{Kpptot} (d), which is compared with the $K^-pn$ formation {spectrum} in Fig.~\ref{Kpptot} (b) calculated with the same optical potential strength.
These spectra have very similar shape qualitatively, but are different 
in the absolute values. We can learn the influences of the elementary 
cross section and the distortion factor to the formation spectra 
from these figures. We have also confirmed that the spectrum of 
the ${\bar K^0}nn$ formation in the $t$($K^-,n$) reaction 
resembles that of $K^-pp$ shown in Fig.~\ref{Kpptot} (a).

\subsection{Semi-exclusive spectra of ${\bar K}NN$ system formation}
\label{sec:4-2-2}
We show in this subsection the
{semi-}exclusive (${\bar K},N$) reactions spectra in coincident with the particle pair emissions due to kaon absorption in ${\bar K}NN\rightarrow MBN$ one-body processes, where $M$ and $B$ indicate the octet mesons and baryons, respectively.
We indicate the spectator nucleon as $N$ in the final state.

\begin{figure*}[htpd]
\includegraphics[width=13.cm,height=8.0cm]{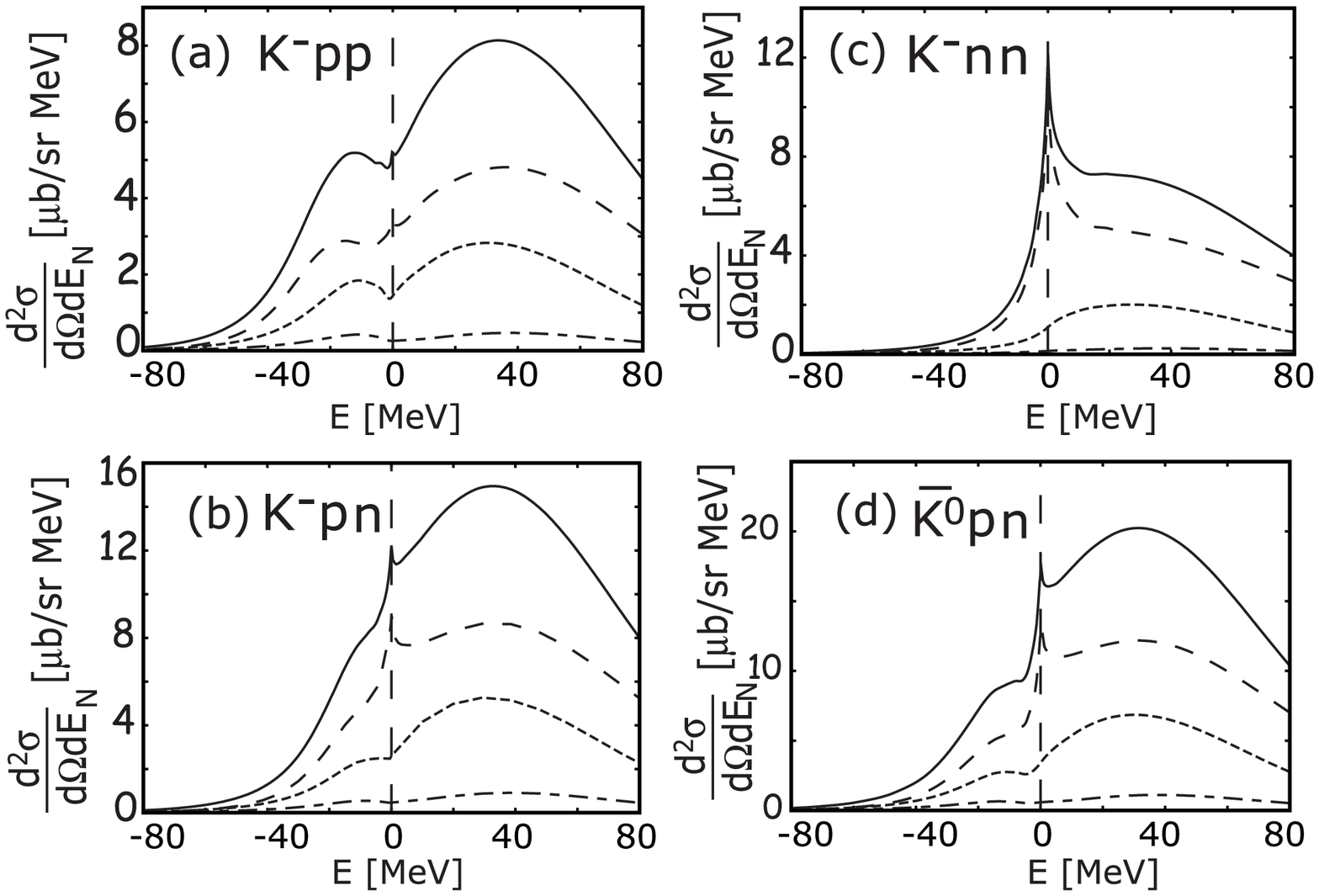}
\caption{\label{conversion}
Conversion parts of the calculated formation spectra of (a)$K^-pp$ in $^3$He($K^-,n$), (b)$K^-pn$ in $^3$He($K^-,p$), (c)$K^-nn$ in $t$($K^-,p$), and (d)${\bar K^0}pn$ in $^3$He($K^-,n$) reactions at $T_{\bar K}=600$ MeV ($P_{\bar K}=976$ MeV/c) are plotted as a function of the real kaon energy at $\theta_N^{\rm lab}=0$ (deg.) for the $s$-wave chiral unitary optical potential.
The horizontal axis $E$ indicates the real kaon energy.
Solid lines show the total conversion spectra.
Dashed, dotted and dot-dashed lines indicate the contributions from kaon $s$, $p$, and $d$ partial waves in the final state, respectively.
The vertical dashed line indicates the kaon production threshold.}
\end{figure*}

First, we consider the conversion part of the formation spectra defined in Eq.~(\ref{eq:6}) with the $S_{\rm con}$ in Eq.~(\ref{eq:9})~\cite{morimatsu85}, which corresponds to the (${\bar K},N$) spectra accompanied by the kaon absorption.
The calculated results are shown in Fig.~\ref{conversion} for the four kinds of the ${\bar K}NN$ systems.
By comparing these results with the total spectra in Fig.~\ref{Kpptot}, we have found that the spectra in the kaon bound region are more clearly seen in the conversion spectra.
\noindent
For example, we can compare Fig.~\ref{conversion} (a) with the Fig.~\ref{Kpptot} (a) for the $K^-pp$ formation.
{Since the total spectra in the kaon positive energy region include the $K^-$ escaping contribution, which is removed by considering the conversion, the conversion spectrum in Fig.~\ref{conversion}(a) has prominent bump structure in the subthreshold region.}
Thus, by measuring the conversion spectra, we can expect to obtain clearer signals in (${\bar K},N$) spectra.

\begin{figure*}[htpd]
\includegraphics[width=13cm,height=8.0cm]{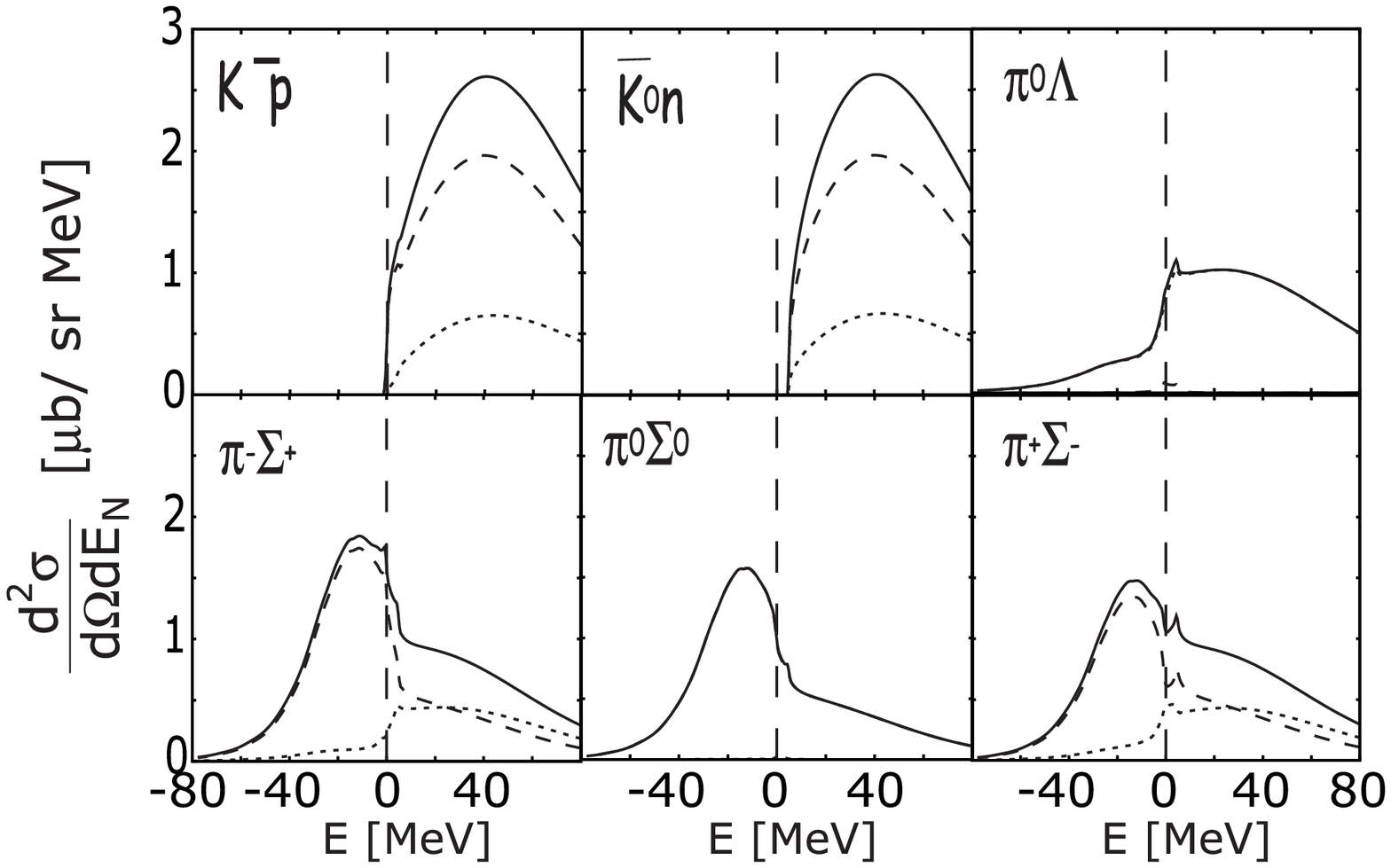}
\caption{\label{conversionKpp}
{Semi-exclusive} spectra for the formation of the $K^-pp$ system in $^3$He($K^-,n$) reactions at $T_{K^-}=600$ MeV ($P_{K^-}=976$ MeV/c) at $\theta_n^{\rm Lab}=0$ (deg.) for the $s$-wave chiral unitary optical potential.
As indicated in the figure, each figure corresponds to the different meson($M$)-baryon($B$) emission channel after one-body kaon absorption process $K^-N\rightarrow MB$.
Solid line shows the total spectrum for each
{semi-}exclusive channel.
Dashed and dotted lines indicate the contributions of ${\bar K}N$ isospin$=0$ and $1$ optical potential, respectively.
The vertical dashed line indicates the kaon production threshold.
}
\end{figure*}

{Next, we proceed the step further. We calculate
{semi-}exclusive spectra 
by dividing the conversion spectra into each particle state as formulated in Eqs. (\ref{Scon_con})~and (\ref{chiralT_con}).}
The results of the
{semi-}exclusive spectra are shown in Fig.~\ref{conversionKpp} for the $K^-pp$ system formation.
As we can see from the figure, by measuring the
{Especially, the peak structures appearing in the subthreshold region in 
the $\pi \Sigma$ emission spectra are prominent. This is because the 
$K^{-}pp$ bound states are driven mainly by the strong attraction 
in the $I=0$ $\bar KN$ channel with the $\Lambda(1405)$
resonance, which decays into the $\pi \Sigma$ state.}
Thus, our calculated results indicate that the $^3$He($K^-,n$) reaction in coincident with the $\pi\Sigma$ emission by $K^-p\rightarrow \pi\Sigma$ one-body kaon absorption is one of the best choice to observe the subthreshold strength of the spectra to obtain the information on the $K^-pp$ state formation.
{In the $K^-p$ and ${\bar K^0}n$ emission channels, we have the contributions only for the quasi-free energy region, since the $\bar KN$ emissions below the threshold are kinematically forbidden.}

\begin{figure*}[htpd]
\includegraphics[width=13.cm,height=5.5cm]{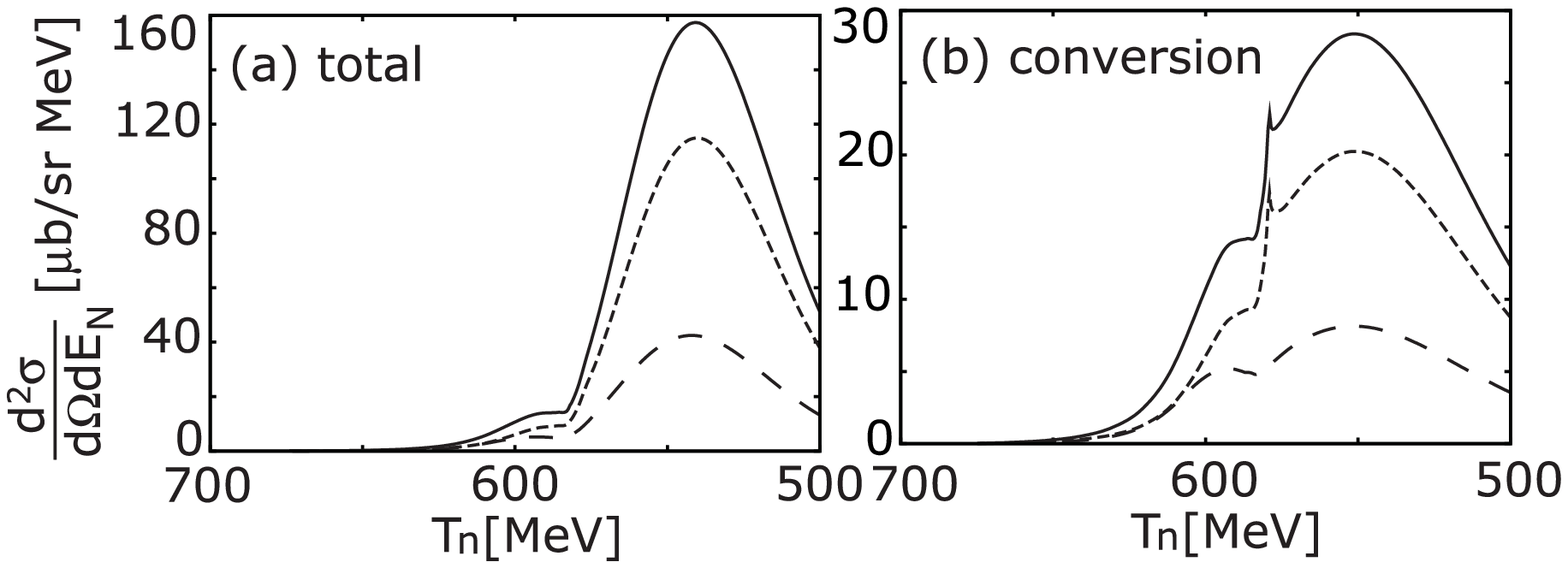}
\caption{\label{Kpp_K0barpnspectra}
Calculated results of $^3$He($K^-,n$) reaction spectra at $T_{K^-}=600$ MeV ($P_{K^-}=976$ MeV/c) including both $K^-pp$ and ${\bar K^0}pn$ in the final states are shown in (a) total spectra and (b) conversion part at $\theta_n^{\rm Lab}=0$ (deg.) for the $s$-wave chiral unitary optical potential.
Dashed and dotted lines indicate the contributions from $K^-pp$ formation and ${\bar K^0}pn$ formation, respectively.
Solid lines are the sum of the both contributions.}
\end{figure*}

In the $^3$He($K^-,n$) reaction, both the $K^-pp$ and ${\bar K^0}pn$ states
are produced and experimentally these states cannot be separated out
in the observation of the formation spectrum.  Thus, two contributions should
be summed up when we compare the calculated spectrum with experimental
observation. In the impulse approximation, the spectra for the $K^-pp$ and ${\bar K^0}pn$ systems are incoherently summed, and the result is shown
in Fig.~\ref{Kpp_K0barpnspectra}. 
In this figure, we plot the spectrum as a function of the emitted neutron energy, which is calculated by including the recoil kinetic energy of the kaon-nucleus system, since the kaon production thresholds are different in the $K^-pp$ and ${\bar K^0}pn$ systems.
From the figure, we find that
the signals of the bound energy region in the conversion spectra become smaller and less clear by including the effects of ${\bar K^0}pn$ formation.
However, still there are certain strength in this region, which could be {observed} in the experiments.


{
For the comparison of the theoretical calculations of the formation spectra
with experimental observations of the $\bar KNN$ system, it is interesting 
to consider
{semi-}exclusive spectra of two-baryon emissions, which are planed 
to be observed in the experiment performed at J-PARC~\cite{J-PARC}. The two-baryon
emissions correspond to the non-mesonic decays of the $\bar KNN$ system,
or equivalently the two-nucleon absorptions of $\bar K$. The non-mesonic 
decay of the kaonic nuclei has been recently discussed in Ref.~ \cite{Sekihara:2009yk}.
The
{semi-}exclusive spectra of the $YN$ emissions will be discussed elsewhere~\cite{yamagata09}.
}

\subsection{Uncertainties of theoretical spectra}
\label{sec:4-3}
\begin{figure}[htpd]
\includegraphics[width=8.6cm,height=6.0cm]{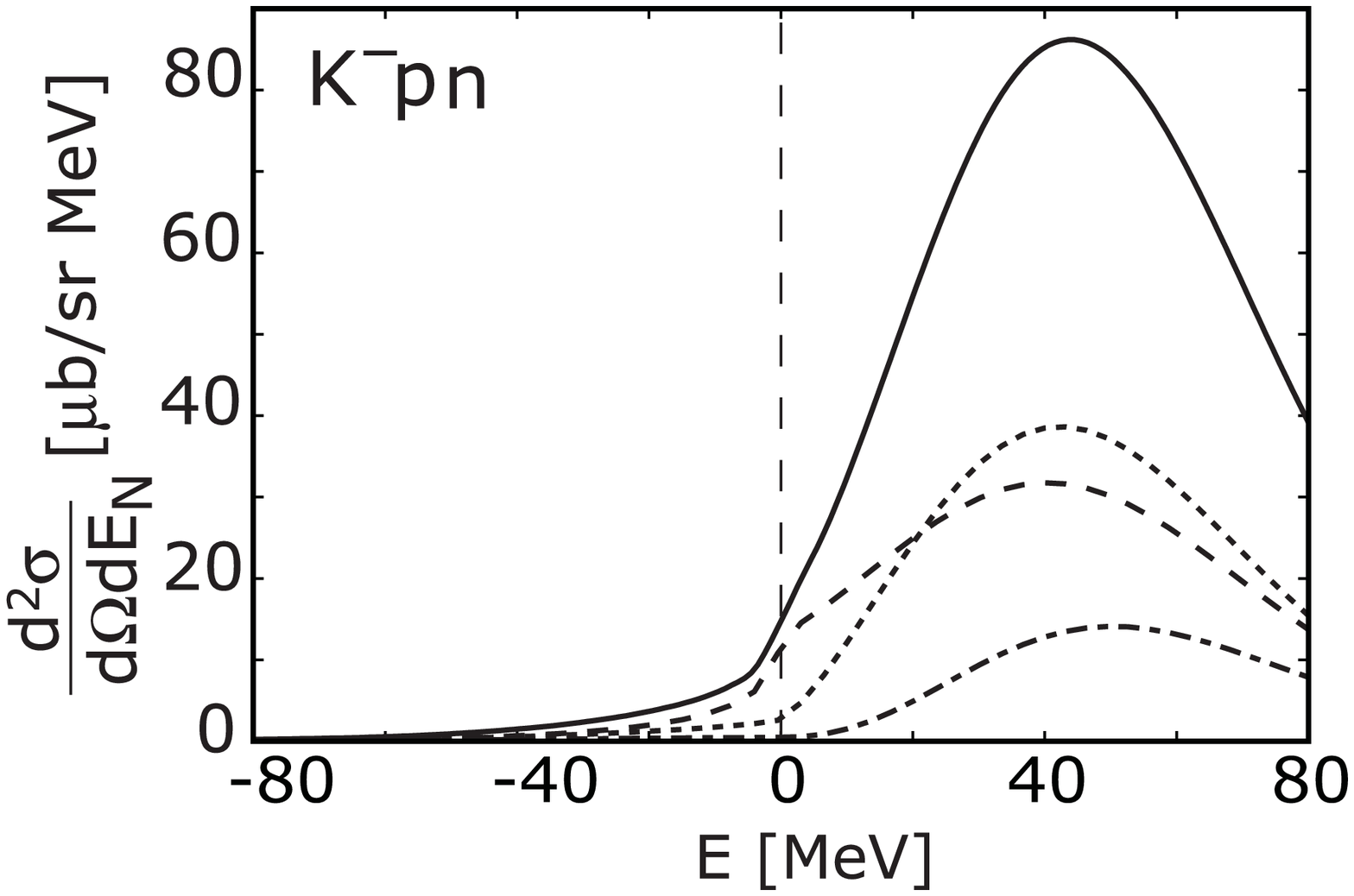}
\caption{\label{Kpn_mediumtot}
Same as Fig.~\ref{Kpptot} (b), except for the $K^--pn$ optical potential used in the calculation.
The $K^-$ selfenergy $\Pi$ in the symmetric nuclear matter with medium effects \cite{ose} is used as the $K^-$-$pn$ optical potential.}
\end{figure}

To see the effects of the medium modifications of the chiral unitary amplitudes, we show in Fig.~\ref{Kpn_mediumtot} the calculated $^3$He($K^-,p$) spectra for the formation of the $K^-$-$pn$ systems using the optical potential shown in Fig.~\ref{Voptmedium} based on the $K^-$ selfenergy evaluated {beyond the low density approximation} in Ref.~\cite{ose} for symmetric nuclear matter.
By comparing Fig.~\ref{Kpn_mediumtot} with Fig.~\ref{Kpptot} (b), we find that the subthreshold behavior in Fig.~\ref{Kpn_mediumtot} is smoother than Fig.~\ref{Kpptot} (b) and does not show the bump structure because of the larger imaginary potential in this energy region.
The selfenergy used to obtain Fig.~\ref{Kpn_mediumtot} includes the higher-order medium effects such as multi-nucleon $K^-$ absorption processes in the matter, although the $K^-$-$pn$ system considered here, we need to evaluate the medium effects up to two-body processes{. Thus, we expect that the effects of the imaginary potential would not be so strong as seen in Fig.~\ref{Kpptot} and that} the medium effects {would} not change the spectra significantly, {but} the signal could be a little unclearer due to the inclusion of two-body absorption.

\begin{figure}[htpb]
\includegraphics[width=8.6cm,height=6.0cm]{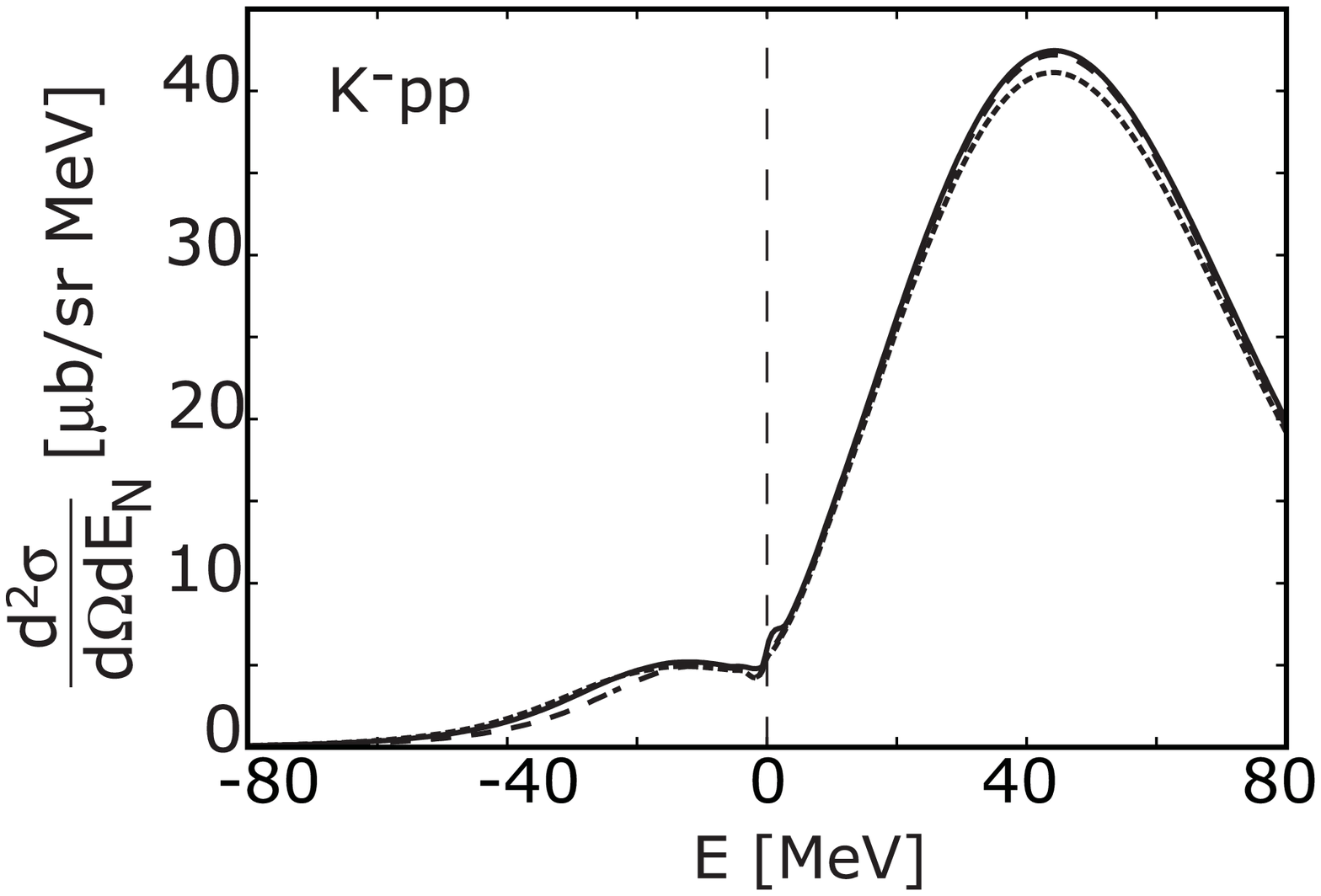}
\caption{\label{rhochange}
Calculated results of $^3$He($K^-,n$) spectra at $T_{K^-}=600$ MeV ($P_{K^-}=976$ MeV/c) for the formation of $K^-pp$ state at $\theta_n^{\rm Lab}=0$ (deg.) with the $s$-wave chiral unitary optical potential. 
Solid line shows the same result as that in Fig.~\ref{Kpptot} (a).
Dashed and dotted lines indicate the results with the $pp$ distributions 
{with $R_{pp}=1.58$ fm and $0.68$ fm, which have 50\% lower and 50\% higher central densities,}
respectively.
See details in text.}
\end{figure}

{
We also study the proton density profile dependence of the $^3$He($K^-,n$) total spectra for the $K^-pp$ formation. In the previous spectrum calculations
shown in Fig.~\ref{Kpptot} to Fig.~\ref{Kpp_K0barpnspectra}, we have used 
the density profile given by the Fermi distribution with the radius 
$R_{pp}=1.01$ fm and the diffuseness $a_{pp}=0.50$ fm. We also calculate 
the formation spectra of the $K^{-}pp$ systems with the density distributions 
which have the central density with $50\%$ lower and $50\%$ higher. 
These density configurations are achieved in the Fermi distributions with
$R_{pp}=1.58$ fm and $0.68$ fm, respectively, which are obtained by 
adjusting the radius parameters with keeping the normalization and 
the diffuseness. The calculated formation spectra with these density 
profiles are shown in Fig.~\ref{rhochange} together with the spectrum 
obtained in the previous calculation.  As seen in the figure,
the spectra are insensitive to the $50\%$ variation of the central density.
We have carried out the same study for the $K^-nn$ formation spectra and confirmed that the spectra are also insensitive to the 
change of the density distribution.
Having also performed the study of the density profile dependence 
of the formation spectra with a different $\bar K$ optical potential,
which is given in Ref.~\cite{Yamazaki:2002uh}, we have obtained the similar  
formation spectra within the variation of the density distributions. 
}

\begin{figure}[b]
\includegraphics[width=8.6cm,height=6.0cm]{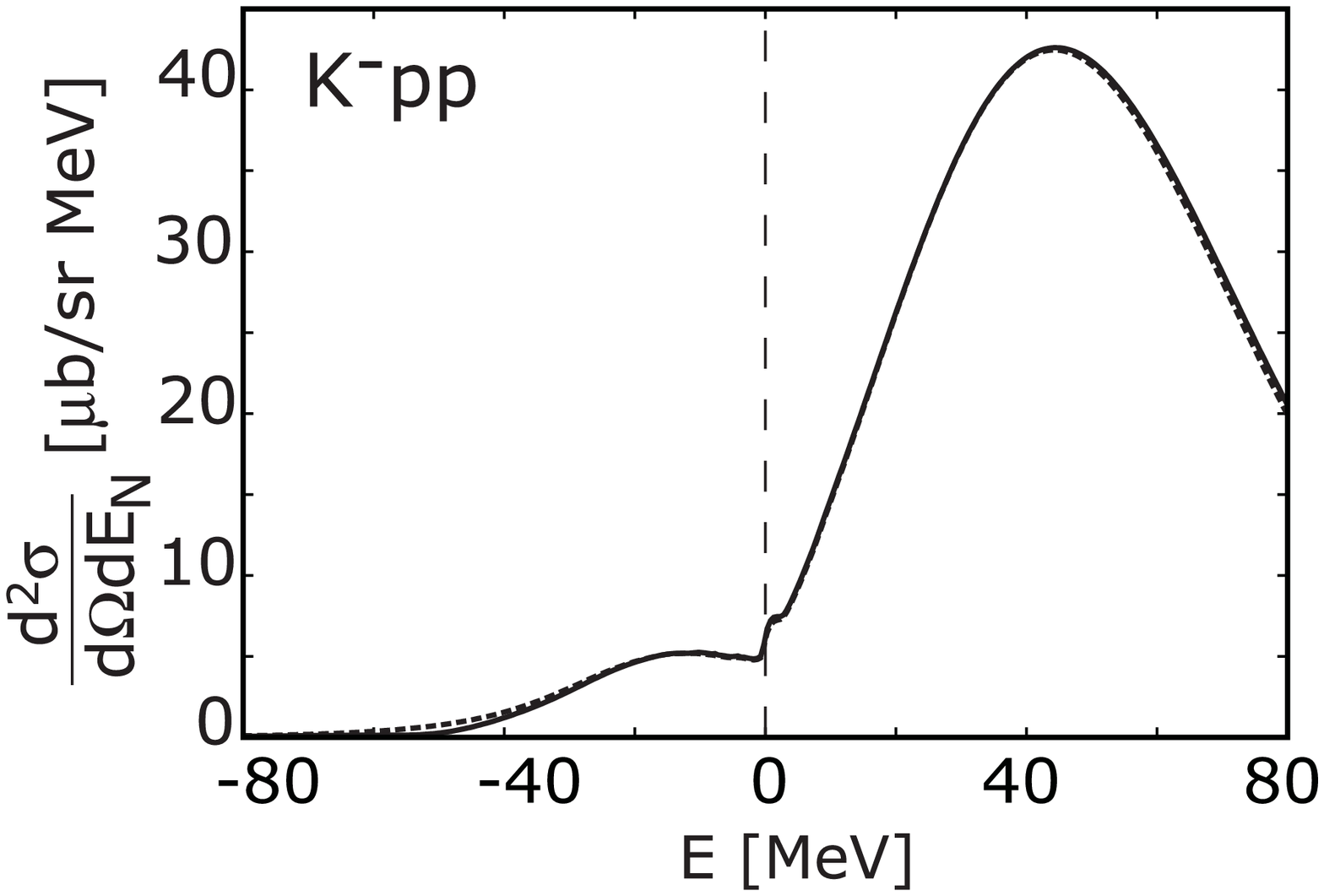}
\caption{\label{spKpptot}
Same as Fig.~\ref{Kpptot} (a), except for the $K^--pp$ optical potential used in the calculation.
Dashed line indicates the result calculated with the $s$-wave chiral unitary optical potential~(Eq.~(\ref{eq:Vopt})), and solid line indicates that with $s$- and $p$-wave optical potential~(Eq.~(\ref{eq:Vopt_p})).
The $p$-wave chiral amplitude calculated in Ref.~\cite{Jido:2002zk} is used to evaluate the optical potential.
}
\end{figure}

In Fig.~\ref{spKpptot}, we show the contribution from $p$-wave optical potential, which takes account of the $\Sigma(1385)$ resonance effects in ${\bar K}N$ channel.
We have used the $p$-wave scattering amplitude calculated in Ref.~\cite{Jido:2002zk} in chiral unitary model and used the optical potential defined in Eq.~(\ref{eq:Vopt_p}).
We found that the contributions from the $p$-wave optical potential is tiny for the total spectra in our present theoretical framework as in Fig.~\ref{spKpptot}.
This results is consistent with that in Ref.~\cite{GarciaRecio:2000nu}, where the $p$-wave potential effects are very small for kaonic atoms.
The effects of the $p$-wave potential and/or $\Sigma(1385)$ resonances are also reported in Refs.~\cite{Dote:2007rk,Dote:2008hw,Wycech:2008wf} in the studies of the structure of the ${\bar K}NN$ systems, where these effects are not negligible for quantitative discussions.

Finally, we would like to add a few comments on the uncertainties of our results.
In our framework, only one body operator is considered as shown in Eq.~(\ref{tau}), and the momentum transfer in the (${\bar K},N$) reaction is shared by the high momentum components of the wavefunctions of a nucleon in the target in the initial state and a kaon in the final state as in Ref.~\cite{Koike:2007iz}.
The many body reaction processes, which we have not evaluated here, exist in reality and have certain contributions to the reaction spectra.
We think that the contributions from the many body processes are structureless and mainly appear in the lower emitted proton energy region {in general}, which corresponds to the quasi-free kaon production region.
Hence, we think the calculated results in the quasi-free kaon production region can be affected by the many body processes~\cite{kishimoto_final}.
{Several processes may have finite contributions to the spectrum in the bound kaon energy region as backgrounds as discussed in Ref.~\cite{gata07}, which will be evaluated quantitatively in future works.}

\subsection{($K^-,p$) spectrum for $^{12}$C target}
\label{sec:4-4}

\begin{figure}[htpd]
\includegraphics[width=8.6cm,height=6.0cm]{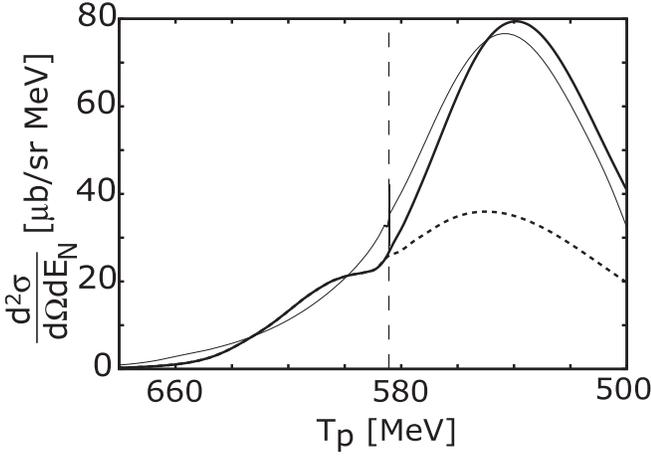}
\caption{\label{fig:12C}
Calculated spectra of the $^{12}$C($K^-,p$)$^{11}$B$\otimes K^-$ reaction at $T_{K^-}=600$ MeV ($P_{K^-}=976$ MeV/c) plotted as a function of the emitted proton energy $T_p$ at $\theta_p^{\rm Lab}=0$ (deg.) for the $s$-wave chiral unitary optical potential.
Thick solid and dotted lines indicate the total and conversion spectra calculated with the free space chiral amplitude, respectively.
Thin solid line indicates the total spectrum calculated with the in-medium chiral amplitude reported in Ref.~\cite{gata06}.
The vertical dashed line indicates the kaon production threshold with the ground state of $^{11}$B.}
\end{figure}

Finally, we show the calculated spectra of the $^{12}$C($K^-,p$)$^{11}$B$\otimes K^-$ reaction at $T_{K^-}=600$ MeV with the chiral unitary amplitude.
In Fig.~\ref{fig:12C}, we show the total and the conversion spectra calculated with the chiral amplitude in free space, together with the total spectra with the amplitude at finite density for comparison~\cite{gata06}.
As we can see from the Fig.~\ref{fig:12C}, the spectra around the threshold becomes smooth and structure-less due to the medium effects to the chiral unitary amplitude.
This is the same tendency which we have observed for the $K^-pn$ systems in Fig.~\ref{Kpptot} (b) and Fig.~\ref{Kpn_mediumtot}.
The medium effects which include the multi-nucleon processes for kaon absorption make it more difficult to observe clear structures in the total spectrum.

\begin{figure*}[htpd]
\includegraphics[width=12.5cm,height=10.0cm]{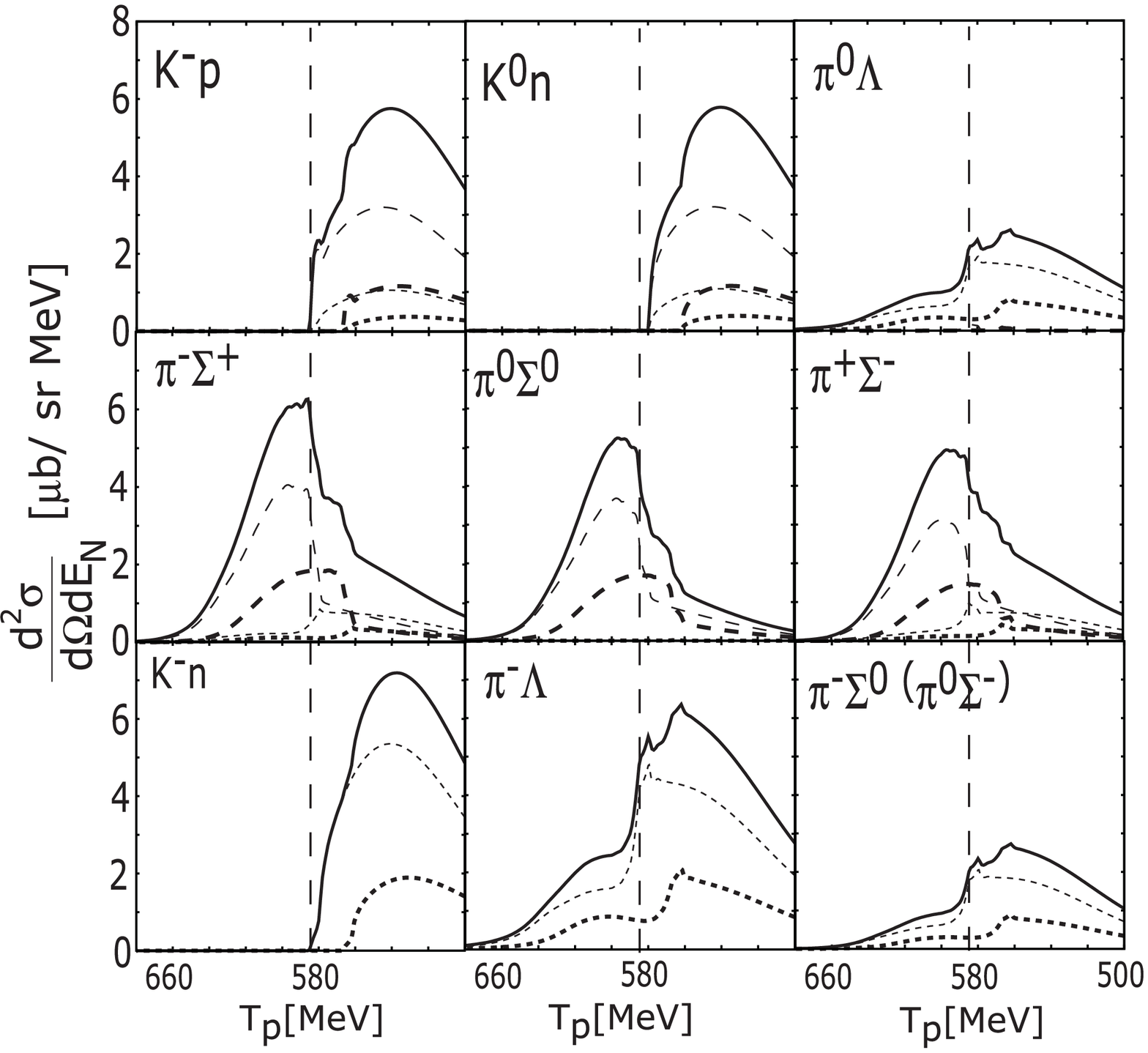}
\caption{\label{fig:12C_conv}
As indicated in the figure, each figure corresponds to the different meson ($M$)-baryon($B$) emission channel after one-body kaon absorption process $K^-N\to MB$.
Due to the isospin symmetry, the spectrum with $\pi^0\Sigma^-$ emission channel is the same as that of $\pi^-\Sigma^0$ channel.
Upper and middle six panels show the emission channels from $K^-+p$ and lower three panels the emission channels from $K^-+n$, respectively.
Thick solid line show the total spectrum for each
{semi-}exclusive channel.
Thick and thin dashed lines indicate the contributions of ${\bar K}N$ isospin=0 with $s_{1/2}$ and $p_{3/2}$ proton hole states, and thick and thin dotted lines indicate those of ${\bar K}N$ isospin=1 with $s_{1/2}$ and $p_{3/2}$ proton hole states, respectively.
There are no isospin=0 contribution in the $K^-+n$ amplitude (lower three panels).
}
\end{figure*}

We show in Fig.~\ref{fig:12C_conv} the
{semi-}exclusive spectra for the $^{12}$C($K^-,p$)$^{11}$B$\otimes K^-$ reaction calculated with the chiral amplitude in free space.
These figures show the ($K^-,p$) spectra with the meson and baryon emissions due to kaon absorption as explained for the $K^-pp$ system case in Fig.~\ref{conversionKpp}.
In the $^{12}$C target case, we have two different features from the $K^-pp$ system, which are the contributions from $p_{3/2}$ proton hole state in $^{11}$B and the kaon absorption by neutron.
In Fig.~\ref{fig:12C_conv}, we show the contributions from different ${\bar K}N$ isospin and proton-hole states separately.
The contributions of kaon absorption by neutron are shown in three lower panels.
We find that the qualitative features of the
{semi-}exclusive spectra are resemble to the $K^-pp$ case and we find again that the $\pi\Sigma$ emission channel from the initial $K^-p$ subsystem have large components in the bound energy region.
We also find that the $\pi^-\Lambda$, $\pi^-\Sigma^0$ and $\pi^0\Sigma^-$ emission channels from $K^-n$ initial subsystem have certain strength in subthreshold kaon energies.
Though we do not include the final state interaction for the emitted meson and baryon after kaon absorption, it would be interesting to explore the
{semi-}exclusive spectra even for larger nuclei such as $^{12}$C since we have possibilities to observe clearer structure in the spectra and to obtain information on subthreshold kaon properties at finite density, which could not be observed in total spectra alone.


\section{\label{sec5}Conclusion}
We have made systematic studies for the formation spectra of the ${\bar K}NN$ and $K^-$-$^{11}$B systems which are accessible by the (${\bar K},N$) reactions.
We have adopted the theoretical ${\bar K}N$ amplitude obtained by the chiral unitary model and formulated the optical potential within the so-called $T\rho$ approximation to calculate the formation spectra systematically.
This theoretical optical potential has the complex energy dependence in contrast to the phenomenological potentials for which we need to assume (or neglect) the energy dependence.
{Although we have used the Green's function with the density dependent optical potential for the ${\bar K}NN$ systems rather than treating them as few-body systems, this simplification makes the spectrum calculation much simpler.}

In order to know the basic features of the light kaonic nuclear systems, we solve the Klein-Gordon equation with the theoretical potential selfconsistently for kaon in proton matter and for ${\bar K}NN$ bound states with assuming the two nucleon density distributions.
From the behaviors of poles in the proton matter, we found that the pole in the kaon bound energy region is connected to one of the poles of $\Lambda(1405)$ in low density limit and not connected to free kaon in vacuum.
The solutions of the Klein-Gordon equation of the ${\bar K}NN$ systems show clearly the differences of the structure of the bound spectra for $K^-pp$, $K^-pn$, and $K^-nn$ states due to the different characters of the interactions.

As for the formation spectra of the light kaonic nuclei, we have made detail analyses along to the line described in Sec.~\ref{sec:intro}.
We would like to emphasis that the
{semi-}exclusive spectra for (${\bar K},N$) reaction coincident with the particle emissions due to kaon absorption are important to obtain the experimental information on the subthreshold kaon properties at finite density.
Especially, in $\pi\Sigma$ emission channels we expect to obtain subthreshold kaon properties and clearer indications of kaon bound states as enhancements in the spectra.
Namely it is possible to obtain the information on the existence of bound states by knowing the strength of the subthreshold spectrum.
On the other hand, it is difficult to know the binding energies and widths quantitatively from the spectra in general for the systems with large width.

Finally, we mention the effects of the mixture of the ${\bar K^0}pn$ formation process to $K^-pp$ formation in the $^3$He($K^-,n$) reaction.
This effect should be evaluated since it is included experimental spectra as a kind of background.
We have reported the calculated results in our theoretical framework in this article.

We believe that our theoretical results are interesting and stimulating for both theorists and experimentalists in this field and help to develop the research of the kaon physics.

\begin{acknowledgments}
This work is partly supported by Grants--in--Aid for scientific research of MonbuKagakusho and Japan Society for the Promotion of Science No. 18-8661 (H.N.), No. 19-2831 (J.Y.), No. 20028004 (D.J.), and No. 20540273 (S.H.).
J. Y. is the Yukawa Fellow and this work is partially supported by Yukawa Memorial Foundation.
This work was partially done under Yukawa International Program for Quark-Hadron Science.
\end{acknowledgments}

\end{document}